\documentclass[aps,prd,nofootinbib,twocolumn,superscriptaddress,preprintnumbers,letterpaper,natbib,nofootinbib]{revtex4-2}

\RequirePackage{arydshln}

\pdfoutput = 1
\usepackage{amssymb}
\usepackage{amsmath}
\usepackage{epsfig}
\usepackage{hyperref}
\usepackage{color}
\usepackage{cleveref}
\usepackage{multirow}
\usepackage{capt-of}
\usepackage{soul}
\usepackage{siunitx}
\usepackage{graphicx}
\usepackage[caption=false]{subfig}
\usepackage{slashed}
\usepackage{tabularx}
\usepackage{enumitem}
\usepackage{multirow}
\usepackage{comment}
\usepackage{url}
\usepackage{xspace}
\usepackage{hhline}

\newcommand{\ord}[1]{\mathcal{O}{(#1)}}
\newcommand{\beq}{\begin{equation}}
\newcommand{\eeq}{\end{equation}}

\newcommand{\bea}{\begin{eqnarray}}
\newcommand{\eea}{\end{eqnarray}}

\makeatletter
\def\simgt{\mathrel{\lower2.5pt\vbox{\lineskip=0pt\baselineskip=0pt
           \hbox{$>$}\hbox{$\sim$}}}}
\def\simlt{\mathrel{\lower2.5pt\vbox{\lineskip=0pt\baselineskip=0pt
           \hbox{$<$}\hbox{$\sim$}}}}
\makeatother

\newcommand{\B}{\mathcal{B}}

\newcommand\psiB{\psi_{\mathcal{B}}}
\newcommand\psiBbar{\bar{\psi}_{\mathcal{B}}}

\newcommand{\vphi}{\varphi}

\newcommand{\Y}{\mathcal{Y}}

\newcommand{\Bri}{\text{Br}\left(B_i^0\rightarrow \psiBbar \, \mathcal{B}_{\rm SM}\right)} 

\begin{document}

\title{Mesogenesis through the Ephemeral Dark Decay of Beauty}

\author{Hooman Davoudiasl}
\email{hooman@bnl.gov}
\affiliation{High Energy Theory Group, Physics Department \\ Brookhaven National Laboratory, Upton, NY 11973, USA}

\author{Rachel Houtz}
\email{rachel.houtz@ufl.edu}
\affiliation{Department of Physics, University of Florida, Gainesville, FL 32611, USA}

\author{Seyda Ipek}
\email{Seyda.Ipek@carleton.ca}
\affiliation{Carleton University  1125 Colonel By Drive, Ottawa, Ontario K1S 5B6, Canada}

\begin{abstract}

{\it Mesogenesis}  provides a path for generating the baryon asymmetry of the Universe, using only the \emph{CP} violation furnished by the Standard Model  in the decay of $B$ mesons.  While this is an intriguing possibility, it is largely constrained by the data on $B$ meson branching fractions into baryons and missing energy carried into the dark sector.  We point out that it is possible to make this branching fraction dominant only in the early Universe, through an ultralight scalar coupled to the dark sector  and the Standard Model leptons. A scenario is examined where the thermal density of muons in the early Universe temporarily lowers the mass of a dark fermion,  allowing for efficient $B$ meson decays. This ``dark" decay channel is shut off later when the muon number density falls, making the scenario compatible with flavor data.  Our model can be consistent with the LHC constraints on color-charged heavy bosons  required to implement  mesogenesis; such states may be discovered in the future runs as their masses  cannot be far above  the current bounds.  We also outline other possible signals, which can arise in future displaced vertex searches, long range force searches, and  observations of neutron star binary mergers. 

\end{abstract}

\maketitle

\section{Introduction}

The mechanism for generation of the observed baryon asymmetry of the Universe (BAU) \cite{pdg} 
\beq
Y_{\mathcal B}\equiv (n_{\mathcal B} - n_{\bar{\mathcal B}})/s = (8.718 \pm 0.004)\times 10^{-11}\,,
\label{eq:YB}
\eeq
remains an open question in particle physics and cosmology.  In the above, $n_{\mathcal B}$ and $n_{\bar{\mathcal B}}$ are the number densities of baryons and antibaryons, respectively; $s$ is the entropy density.  Quite generally, it is expected that whatever mechanism generated the BAU must satisfy the Sakharov conditions \cite{Sakharov:1967dj}, one of which is the presence of \emph{CP} violation.  While the Standard Model (SM) contains \emph{CP} violating interactions, it is widely believed that it is by far not at sufficient levels~\cite{Gavela:1993ts, Gavela:1994ds, Gavela:1994dt, Huet:1994jb}.  This circumstance has been one of the main motivations to postulate and look for new \emph{CP} violating physics.

A counter example to the above conclusion was proposed in Ref.~\cite{Elor:2024cea}, where the generation of the BAU was sourced by the \emph{CP} violation present in the SM, specifically in certain $B$ meson decays in the early Universe.\footnote{For other examples of utilizing the CKM phase for baryogenesis by dynamical effects in the early Universe, see \cite{
Berkooz:2004kx,Perez:2005yx,Bruggisser:2017lhc, Baldes_2016,
von_Harling_2017, 
Baldes_2018,Bruggisser_2018, bruggisser2018electroweakphasetransitionbaryogenesis,Fuyuto:2017ewj}.} The model in Ref.~\cite{Elor:2024cea} is a mesogenesis model \cite{Elahi:2021jia, Elor:2018twp, Alonso-Alvarez:2019fym, Nelson:2019fln} modified to overcome exotic $B$-meson  branching fraction constraints as well as LHC constraints on color-triplet scalars in order to be able to utilize the measured \emph{CP} violation in the $B$ meson sector. In mesogenesis, a color-triplet scalar particle mediates interactions between $b$-quarks and the dark sector. (The new interactions are not necessarily baryon number violating since the dark sector can also carry baryon number. However, baryon number in the visible sector can still be \emph{violated} while being balanced by an equal and opposite asymmetry in the dark sector.) In Ref.~\cite{Elor:2024cea}, it is proposed that this mediator go  through a mass change in the early Universe, starting with $M_{\mathcal{Y}}=O(100~{\rm GeV})$ before the generation of the BAU and ending with an $O({\rm TeV})$ mass, currently. These different masses allow for a large branching fraction for the $\B$-violating $B$-meson decays in the early Universe while being safe from constraints today. 

The ``morphing" of the mediator mass was achieved through generation of domain walls in which another scalar field acquires near-degenerate vacuum expectation values. In half of the domains, the mediator is lighter and the BAU is copiously produced. However, the domains with the heavier mediator is at the global minimum of the scalar potential and eventually the domain walls annihilate.  This intriguing domain wall mechanism, unfortunately, might suffer from constraints from big bang nucleosynthesis (BBN)~\cite{Bagherian:2025puf}. Even though the domain wall annihilation happens before BBN, at temperatures $O(10~{\rm MeV})$, the resulting baryon asymmetry, which is only correctly produced in certain domains, might not be able to homogenize before BBN and this inhomogeneity could lead to incorrect light element abundances. 

In this paper,  we discuss a different mechanism that will allow for a significant exotic $B$-meson branching fraction in the early Universe while being safe from such constraints today. Instead of morphing of the mediator mass, here we focus on the dark sector fermion,\footnote{A similar approach was pursued very recently in Ref.~\cite{Baruch:2026iwn} using different mechanisms for changing the $\psi_B$ mass.} $\psiB$, into which the $B$ meson decays, together with SM quarks [see \cref{eq:UVmodel}, below]. We follow Refs.~\cite{Batell:2021ofv, Croon:2020ntf} to envision a scenario in which coupling an ultralight scalar to SM charged leptons lowers the mass of the dark fermion in the early Universe, compared to today.

In what follows, we will first briefly outline the basic ingredients of mesogenesis.  Next, we will provide a model that makes the key $B$ meson decays that involve the dark sector fermion $\psiB$ dominant while the Universe was at temperatures of $\ord{\rm 10~MeV}$, but kinematically forbidden at later times.  This is achieved by making the $\psiB$ mass dependent on an ultralight scalar that can be sourced by charged leptons in the early Universe.   We will also discuss how the needed effective operators can be generated by certain color charged scalars that can evade current LHC bounds, but may be expected to be probed in the coming years. Finally, we will outline some of the potential signals of our scenario that may arise in laboratory and astrophysical settings.

\section{Mesogenesis: a brief outline} \label{sec:MesoSummary}

We start with a brief summary of the mesogenesis mechanism~\cite{Elahi:2021jia, Elor:2018twp, Alonso-Alvarez:2019fym, Nelson:2019fln}. In general terms, mesogenesis can be described by dim-6 effective operators between SM quarks and a new dark fermion $\psiB$ carrying antibaryon number, $\B=-1$, giving rise to a ($\B$-conserving) effective operator: 
\bea
\mathcal{O}_{d_k, u_i d_j} = \mathcal{C}_{d_k, u_i d_j} \epsilon_{\alpha \beta \gamma} (\psiBbar d_k^\alpha ) (\bar{d}_j^{c \beta} u_i^\gamma)  \,,
\label{eq:EFTop}
\eea
where $i,j$ and $\alpha, \, \beta, \, \gamma$ are flavor and color indices, respectively. Note that in order to kinematically forbid proton decay through the $\psiB u dd$ interaction, we require all dark sector baryons to have a mass $\gtrsim 1 \, \text{GeV}$ at any given time.

A straightforward UV model that gives rise to \cref{eq:EFTop} is one with a (heavy) triplet scalar $\Y$ carrying  electric charge $-1/3$ and $\B=-2/3$. ($\Y$ can be identified with a squark~\cite{Alonso-Alvarez:2019fym}.) Then, we get the following Lagrangian terms:
\bea
\hspace{-0.18in} \mathcal{L}_{\Y} =  -\sum_{i,j} y_{u_i d_j} \Y^\star \bar{u}_{iR} d_{jR}^c - \sum_k y_{\psi d_k} \psiBbar \Y d_{kR}^c  + \text{h.c.}
\label{eq:UVmodel}
\eea
Of course, just as squarks, any color-triplet scalar is typically     constrained by the LHC to be well above the weak scale $\sim 100$~GeV. Let us first explain the generation of the baryon asymmetry. Afterward, we will discuss the collider and $B$-physics constraints which will then guide us in choosing which specific operators to use among  $\mathcal{O}_{d_k, u_i d_j}$.

The model first requires a scalar particle $\Phi$ that decays at a temperature below the QCD scale, $T_d < T_{\rm QCD}\sim 100$~MeV, creating out-of-equilibrium $B$ mesons. $B$ mesons oscillate, giving rise to \emph{CP} violation when they decay. The effective operators in \cref{eq:EFTop} facilitate the exotic decay $B^0_{d,s} \rightarrow \bar{\psi}_{\mathcal{B}} \, \mathcal{B}_{\rm SM}$, into the dark antibaryon and a SM baryon $\mathcal{B}_{\rm SM}$. These decays generate an equal and opposite baryon asymmetry between the dark and SM sectors. (Afterward, $\psiB$ decays into the stable dark matter (DM). This is discussed in \cref{sec:psievo}.) The resulting baryon asymmetry is calculated to be~\cite{Elor:2018twp, Alonso-Alvarez:2019fym}:

\bea
\!\!\!\! Y_{\mathcal{B}} \simeq 5 \times 10^{-5} \, \sum_{i=d,s} \bigl[ \Bri  A_{sl}^i \bigr]  \alpha_i (T_d) \,,
\label{eq:YBmeso}
\eea
where
\begin{subequations}\label{eq:SMCPV}
\begin{align}
&A_{sl}^d|_{\rm SM} = (-4.7 \pm 0.4 ) \times 10^{-4}  \,, \label{eq:aSMd}\\
& A_{sl}^s|_{\rm SM} = (2.1 \pm 0.2) \times 10^{-5}\,, \label{eq:aSMs}
\end{align}
\end{subequations}
are the semileptonic asymmetries in $B^0$ decays generated by the SM \emph{CP} violation \cite{Lenz:2019lvd}. The factor $\alpha_i(T_d)\in [0,1]$ accounts for decoherence effects and depends on the temperature at which $B$ mesons were created through the scalar $\Phi$ decays. It is calculated by solving the Boltzmann equations for the $B$-meson density matrix in the early Universe~\cite{Elor:2018twp}.

From \cref{eq:YB} and \cref{eq:YBmeso}, it is obvious that we need $\Bri \sim 1$ if we want to use the SM \emph{CP} violation in \cref{eq:SMCPV}. However, recast LEP constraints on the Wilson coefficients $\mathcal{C}_{d_k, u_i d_j}$ \cite{Alonso-Alvarez:2021qfd} and dedicated mesogenesis searches by Belle-II \cite{Belle:2021gmc} and BaBar \cite{BaBar:2023rer} collaborations put stringent limits on these branching fractions, $\Bri \lesssim 10^{-4}$. (See Table I-II in Ref.~\cite{Elor:2024cea} for a summary of all available constraints.)

One way to utilize the SM \emph{CP} violation in mesogenesis while still being safe from branching fraction constraints is to only allow the new $B$ meson decays in the early Universe while shutting them off today. In Ref.~\cite{Elor:2024cea} this was achieved by changing the mass of the color-triplet scalar mediator $\Y$ between $T\sim O(10~{\rm MeV})$ and now. It was shown that, with 
\begin{align}
M_\Y \lesssim 620~{\rm GeV} \sqrt{y_{\psi s}y_{cb}}~~~{\rm (successful~mesogenesis)}\,, \label{eq:MYscb}
\end{align}
the operator $O_{s,cb}$ can generate the correct BAU using only the SM \emph{CP} violation when only $B_s$ mesons are the source~\cite{Nelson:2019fln}; see fig.~1 of~\cite{Elor:2024cea}. We will focus on this operator as it allows for the heaviest $\Y$ given a set of Yukawa couplings. (All other operators require much lighter $\Y$ masses to generate the BAU.) Below, we discuss how such a light $SU(3)$-triplet scalar could be allowed. 
\medskip

\textbf{Collider constraints and operator selection.} In the following section, we describe a model in which $\psiB$ mass changes from a few GeV in the early Universe to $\gtrsim 10$ GeV today. With this final mass, the $B$ meson decays into $\psiB$ final states are not allowed. Hence, the previously mentioned Belle-II \cite{Belle:2021gmc} and BaBar \cite{BaBar:2023rer} constraints do not apply. 

\begin{figure}[t]
  \centering
    \includegraphics[width=\columnwidth]{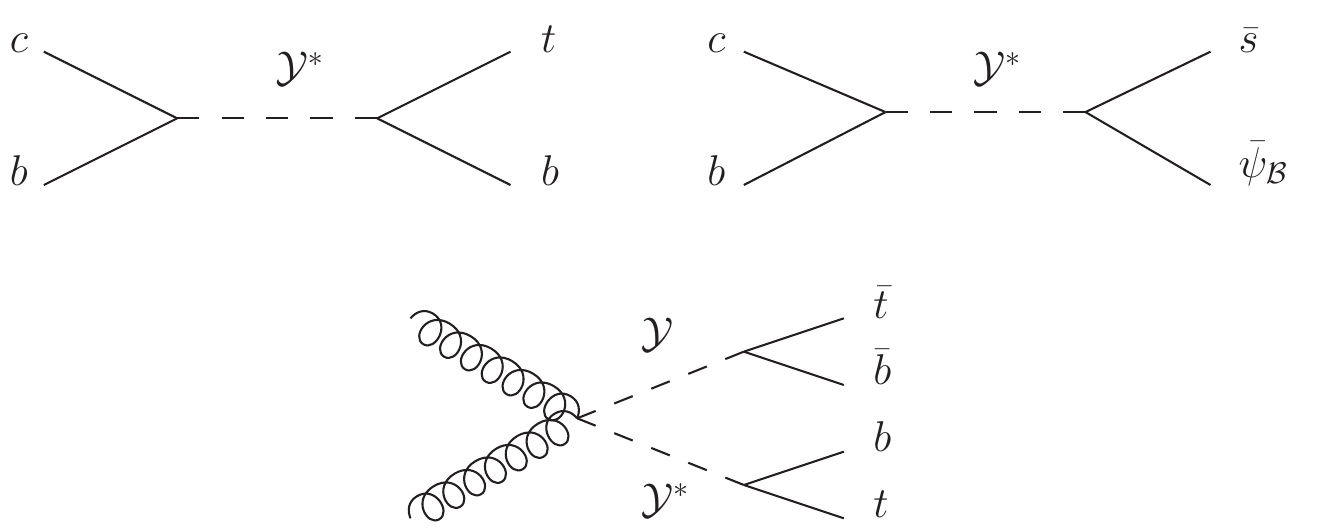}
    \caption{Some production and decay channels for the $SU(3)$-triplet scalar mediator $\mathcal{Y}$ at the LHC.}
\label{fig:channels}
\end{figure}

In Ref.~\cite{Alonso-Alvarez:2021qfd}, constraints on the mediator mass $M_\Y$ and the coupling constants $y_{u_i d_j}, y_{\psi d_k}$ were calculated for single $\Y$ production followed by $\Y$ decays to either multijet or jets + missing transverse energy (MET) final states, assuming a single operator, of the form $\Y^\ast \bar{u}_i d^c_j$, is turned on at a time. (See \cref{fig:channels} for example channels.) Given that the model produces the baryon asymmetry through $B$ meson decays, the analysis only included operators that involve the $b$ quark. A more recent (effective field theory) analysis was carried out in in Ref.~\cite{Hiller:2026osz}. For the $O_{s,cb}$ operator mentioned above, the LHC constraints on jets+MET on single $\Y$ channels translate to 
\begin{align}
   M_\Y \geq 2.4\,{\rm TeV} \sqrt{y_{\psi s}y_{cb}}~~~~({\rm bjet+MET}) \,, 
\end{align}
which is in clear conflict with \cref{eq:MYscb}. ($O_{s,cb}$ is $\mathcal{O}_{232}^{uddN}$ in \cite{Hiller:2026osz}.) The above limit applies to $m_\psi \lesssim 7.5$ GeV, in which case $\psiB$ is collider stable and considered as missing energy. If $\psiB$ is heavier, it can decay into hadron final states with a width approximately given by~\cite{Hiller:2026osz}
\begin{align}
\Gamma(\psiB\to cbs) &\approx \frac{(y_{\psi s}y_{cb})^2\,m_\psi^5}{1024 \pi^3 M_\Y^4}\\ \label{eq:Gammapsi}
&\simeq2\times 10^{-11}~{\rm GeV}(\,y_{\psi s}y_{cb})^2 \notag \\
&\quad\quad\times\left(\frac{m_\psi}{10~{\rm GeV}}\right)^5\left(\frac{600~{\rm GeV}}{M_\Y}\right)^4\,. \notag
\end{align}
The lack of missing energy relaxes the constraint to $M_\Y \gtrsim 1\,{\rm TeV} \sqrt{y_{\psi s}y_{cb}}$. For $7.5~{\rm GeV}\lesssim m_\psi\lesssim 10$ GeV, displaced vertex searches are projected to be sensitive to such final states. Above $m_\psi =10$ GeV, multijet constraints give\footnote{We take this limit from $O_{223}^{uddN}$ as shown in fig.~8 of \cite{Hiller:2026osz}. The operator of interest, $O_{232}^{uddN}$, is expected to be less constrained.}   
\begin{align}
   M_\Y \gtrsim 900\,{\rm GeV} \sqrt{y_{\psi s}y_{cb}}~~~~({\rm multijet}) \,, 
\end{align}
which is still in conflict with \cref{eq:MYscb}.

It is reasonable that two or more of the operators in \cref{eq:EFTop} will be present in a UV model. In that case, there are more stringent constraints coming from $\Delta F= 2$ processes that affect the mixing parameters in neutral meson systems. However, these only apply to different operators that couple to the same quark, e.g. $|y_{td}y^\ast_{ts}|< 3\times 10^{-4}(M_\Y/1.5\,{\rm TeV})$ coming from the kaon mixing parameter  $\epsilon_K$ [see eq. (43) in Ref.~\cite{Alonso-Alvarez:2021qfd}].

With these obstacles  in mind, we use the following Lagrangian,
\begin{align}
    \mathcal{L}_{\rm new} = - y_{\psi s} \Y\,\bar{\psi}_\B s^c_R -y_{cb} \Y^\ast \bar{c}_R b^c_R - y_{td_i} \Y^\ast \bar{t}_R d^c_{iR}  + {\rm h.c.}
\end{align}
for $d_i = d,s,b$. We first note that the addition of top quark interactions do not alter the baryogenesis scenario even with the $\Y \bar{t}b$ operator since the $B$ mesons cannot kinematically decay to a top quark and chain decays would be suppressed by $(m_b/m_t)^2$. In order to relax the collider constraints, we require that $y_{td_i}\gg y_{cb}, y_{\psi s}$. For example, choosing $y_{td_i}\simeq 5 y_{cb}$ can lower the branching fraction of $\Y \to cb$ decays to less than $5\%$, relaxing the relevant constraints from this channel enough to allow for $M_\Y \sim 600$ GeV.\footnote{In order to keep the same exclusion strength, assuming the $s$-channel contributions corresponding to a resonance dominate, a rough estimate is $({\rm Br}\times \sigma)_{900~{\rm GeV}}=({\rm Br}\times \sigma)_{600~{\rm GeV}}$, where $\sigma \propto {\rm PDF}(c){\rm PDF}(b)/M_\Y^4$. A conservative estimate is that the relevant PDFs roughly double going from 900 GeV to 600 GeV. Hence, the new branching fraction to the $cb$ final states need to be ${\rm Br}\lesssim (600~{\rm GeV}/900~{\rm GeV})^4/4\sim 0.05$.}  Note that allowing for all three channels, $\bar{t}d, \bar{t}s,\bar{t}b$ suffers from the flavor constraints from $\Delta F=2$ processes described above. Hence, only one such coupling can be large and the rest needs to be suppressed.

When the dominant interaction is through the $\Y \bar{t} b$ operator, its contribution to single production of $\Y$ is heavily suppressed due to small top quark PDFs. Although in our case $\Y$ can still be produced through the $c\,b$ channel, the dominant final state is now multijets, instead of jets+MET, which are much less constrained.  Of course, as an $SU(3)$-triplet scalar, $\Y$ can be pair-produced through its gluon couplings. There are several LHC searches on pair production of stops with RPV decays. The most relevant one for this case is a CMS search \cite{CMS:2021knz} that excludes stop masses up to 670~GeV (for 100\% branching fractions into top final states). (See \cite{Choudhury:2024ggy} for a recent review of RPV searches.) 

We also note that by adding a large coupling constant, $y_{td_i}\sim 5$, we are effectively broadening the $\Y$ width. As the width of the resonance increases, the background (in a collider search) that one integrates over also increases, hence lowering the signal-to-background ratio. Hence, we expect that the above constraint could relax to $M_\Y \simeq 620$~GeV from $670~$GeV. In Ref.~\cite{CMS:2018mgb} broad and narrow resonance scenarios were discussed for particles with masses $O({\rm TeV})$ with widths up to $\Gamma/M = 30\%$, showing that the background could increase by an order of magnitude from a narrow resonance scenario. The effect is less prominent for smaller masses. It would be interesting to do a dedicated LHC analysis for this particular scenario, with large Yukawas, which we leave for future work.

\section{A model for cosmological  evolution of dark baryon mass} \label{sec:psievo}

In this section, we provide a simple model for making the mass of the dark sector fermion $\psiB$ dependent on the ambient SM particle density.  In particular, we will make $m_\psi$ dependent on the number density of charged leptons $\ell=e,\mu,\tau$.  Given the typical temperatures for mesogenesis  $T_{\rm M} \sim \ord{\rm 10~MeV}$, this is mostly achieved via the $e^\pm$ and $\mu^\pm$ thermal populations.  To do this, we introduce an  ultralight scalar $\vphi$ of mass $m_\vphi$ that couples to $\psiB$ and charged leptons  
\beq
(m_\psi^0 + g_\psi\, \varphi) \psiBbar \psiB + \sum_{\ell=e,\mu,\tau} g_\ell\, \varphi \, \bar \ell \ell \,,
\label{phi-Yukawa}
\eeq
where $g_\psi$ and $g_\ell$ are the couplings of $\vphi$ to $\psiB$ and $\ell$, respectively. Here, $m_\psi^0$ is the mass of $\psiB$ in vacuum, corresponding to $\vphi=0$.  Given the above setup, the mass of $\psiB$ would have a dependence on $\vphi$ via
\beq
m_\psi = m_\psi^0 + g_\psi \, \vphi\,.
\label{mpsi}
\eeq

    We would like to lower the mass of $\psiB$ for $T_{\rm M}\sim 10 $ MeV so that $\Bri$ can proceed effectively during mesogenesis, but the mass later reaches its larger vacuum value,  shutting off these decays after mesogenesis is over. In order for mesogenesis to proceed via our proposed operator, a prompt $B_s \to \psi_{\mathcal{B}}\,\Omega_c^0$ decay must occur. This can only happen when $m_\psi < m_{B_s} -  m_{\Omega_c^0} = 2.7$ GeV. We also require a prompt decay of $\psi_\mathcal{B}$ to dark sector particles $\phi,$ $\xi$ to shut off back-reactions and possible baryon number washout, $m_\psi > m_\phi + m_\xi \gtrsim 1.5$ GeV~\cite{Elor:2018twp}. 
    To safely accommodate these requirements, we conservatively require $m_\psi< 2.5$ GeV during mesogenesis, and we choose a benchmark such that $m_\psi > 2$ GeV for its entire evolution. 

To achieve a reduction of the $\psiB$ mass, we assume $g_\psi,\, g_\ell>0$.  We note that  $\psiB$ eventually decays to DM particles---a scalar $\phi$ and a fermion $\xi$---via an interaction 
\beq
y_\psi\, \phi\, \bar{\psi}_\mathcal{B}\, \xi + {\rm h.c.}
\label{psi-DM}
\eeq
Through these interactions, $\psiB$ can decay to the dark sector faster than washout processes that would repopulate $B$ mesons. To see this, note that the Hubble rate is given by $H(T) = \sqrt{8\pi^3 g_*(T)/90}\, T^2/M_{\rm P}$, where $g_*(T)$ is the number of relativistic degrees of freedom at temperature $T$ and $M_{\rm P}\approx 1.22 \times 10^{19}$~GeV is the Planck mass. $T\sim 10$~MeV corresponds to timescales of $\ord{\rm 0.01~s}$.  The decay width for $\psi\to \xi \phi$, ignoring final state masses, is roughly given by 
\beq
\Gamma(\psiB \to \xi \phi)\approx \frac{y_\psi^2 m_\psi}{32 \pi}\,.
\label{eq:Gammapsitoxiphi}
\eeq
Hence, for a $\psiB$ of mass $\sim$ a few GeV, it only needs to couple with a strength $y_\psi \gg  10^{-10}$ to DM final states in order to decay fast on the relevant cosmological time scales. For $y_\psi \lesssim 10^{-5}$, the prompt cosmological decay of $\psiB$ into DM is compatible with its late time decay dominantly into hadrons.

The initial value of $\vphi$ is sourced by the lepton thermal population and once the Hubble rate falls roughly below $m_\vphi/3$, the scalar begins to oscillate and its amplitude  redshifts as $T^{3/2}$. As $\vphi$ changes, the mass of the dark fermion $\psiB$ also changes. Since we require that $m_\psi$ stays small until $T\lesssim 10$ MeV, we need the $\vphi$ oscillations to be delayed until after mesogenesis as well. Therefore, the mass of $\vphi$ should generally not be much heavier than $m_\vphi\sim O(10^{-13}~{\rm eV})$, making $\vphi$ an  ultralight scalar, with a Compton wavelength exceeding $\sim 10^3$~km. For lighter $\vphi$, the oscillations start later and hence the increase in $m_\psi$ happens later as well. This might result in a period of time where $\psiB$ becomes lighter than the proton allowing proton decay. This effect can give a lower bound on the $\vphi$ mass. To be conservative, we will take $m_\vphi\sim O(10^{-13}~{\rm eV})$ as a safe benchmark value.

Thermal evolution of $\vphi$ is governed by its coupling to particles in the primordial plasma.  In our case, these are charged leptons $\ell$.  In the regime where $\ell$ is relativistic, the evolution of $\vphi$ is given by (see, for example, Ref.~\cite{Croon:2022gwq}) 
\beq
\frac{T^6}{4 M_*^2}\, \frac{d^2\varphi}{dT^2} + 
\left(m_\varphi^2 + \frac{g_\ell^2}{6}\,T^2 \right) \varphi + \frac{g_\ell}{6}\, m_\ell T^2 = 0\,,
\label{vphi-evol-rel}
\eeq
where
\beq
M_* \equiv \frac{M_P}{2}
\sqrt{\frac{90}{8 \pi^3 g_*(T)}}\, .
\label{xi}
\eeq
The corresponding equation for a nonrelativistic lepton $\ell$ is \cite{Croon:2022gwq}
\beq
\frac{T^6}{4 M_*^2}\,\frac{d^2\varphi}{dT^2}   + 
\left[m_\vphi^2 - \frac{g_\ell^2 n_{0\ell}}{T}\left(1-\frac{3 T}{2 m_\ell}\right)
\right] \vphi + g_\ell \, n_{0\ell} = 0\,,
\label{vphi-evol-nonrel}
\eeq
where 
\beq
n_{0\ell} = g\, \left(\frac{m_\ell T}{2\pi}\right)^{3/2} e^{-m_\ell/T}\,,
\label{n0X}
\eeq
with  $g=4$ for a Dirac fermion and its antifermion conjugate ($\ell,\bar\ell)$. Technically, in these equations of motion, $\psiB$ interactions and decays could also be included. However, as discussed above, $\psiB$ decays to DM particles very fast compared to the other time scales. After these decays, there will be a nonthermal density of $\xi$ (and $\phi$) particles which can also shift the $\varphi$ field value through loop interactions. These effects are also negligible; see discussions around \cref{eq:nB,eq:gpsigxi}.

We will take $m_\vphi = 9\times 10^{-14}$~eV as a reference value which suffices to show the feasibility of the mechanism.  We would like $m^0_\psi + g_\psi \langle\vphi\rangle$ to be less than a few GeV during mesogenesis, but larger than $\sim $5 GeV today in order to shut off two-body $B$ meson decays into an on-shell $\psiB$ and an SM baryon. 

A $\vphi$ coupling to muons of $g_\mu = 10^{-24}$ is  unconstrained by about 5 orders of magnitude after  accounting for loop processes that can couple $\vphi$ to electrons and protons~\footnote{The two-loop diagram that couples $\varphi$ to a proton has been recently explicitly calculated in \cite{Fiorillo:2026wso}, which yields $g_p=\alpha^2 g_\mu/2 \sim 3\times 10^{-5} g_\mu$. The current upper bound on proton coupling to a long range force is $g_p\sim 10^{-24}$~\cite{Fayet:2017pdp,MICROSCOPE:2022doy}.} \cite{Berge:2017ovy,Fayet:2017pdp,MICROSCOPE:2022doy,Davoudiasl:2018ltz, Fiorillo:2026wso}.  Even though the muons are quite Boltzmann suppressed at $T\sim T_{\rm M}$, we still find a relatively large background of $\vphi$ that is sourced by them.
\begin{figure}[t]
  \centering
    \includegraphics[width=\columnwidth]{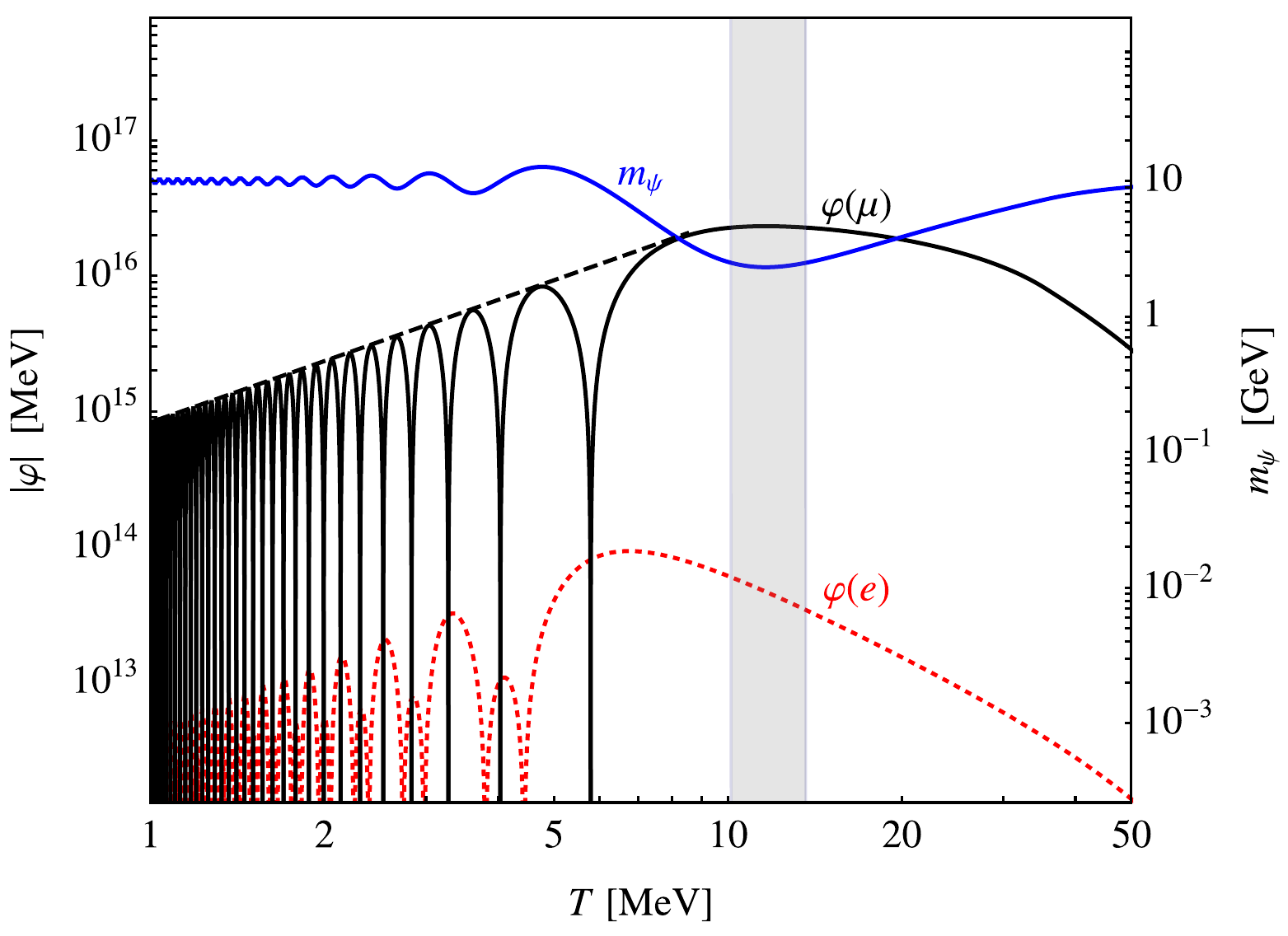}
    \caption{Evolution of the absolute value of $\vphi$ via muon interactions only (black/solid curve, in MeV) and electron interactions only (red/dotted) with temperature (in MeV). We also show  the variation of $m_\psi$ (blue curve, in GeV) with temperature. 
    We have set $g_\psi = 3.3\times10^{-13}$, $g_\mu=10^{-24}$, $g_e = 10^{-25}$, $m_\vphi=9\times10^{-14}$ eV, $m_{\psi}^0 = 10$ GeV. The dashed line is the envelope of the $\vphi$ amplitude that redshifts like $T^{3/2}$, as expected for a component of cold dark matter.  The gray shaded region marks the interval where mesogenesis can take place, corresponding to the regime where a $B$ meson can decay into $\psiB$ on-shell.}
\label{fig:phi-evol}
\end{figure}

As can be seen in \cref{fig:phi-evol}, at $T\sim T_{\rm M}$ we have $\vphi \sim 10^{16}$~MeV.  Hence, if we want to change the mass of $\psiB$ by $\ord{\rm 10^3~ MeV}$, we need to choose $g_\psi \sim 10^{-13}$.  Note that with our reference choices above, $\vphi$ starts to oscillate around $T_1 \approx 4.8$~MeV (first peak in \cref{fig:phi-evol}) with an amplitude of $\vphi \approx 8.4\times  10^{15}$~MeV, corresponding to an energy density
\beq
\rho_\vphi(T_1) = \frac{1}{2}
m_\vphi^2 \vphi^2 \approx  2.8\times 10^{17}~\text{eV}^4, 
\label{rho-phi}
\eeq
which will consequently redshift as $T^3$.  Near the matter-radiation equality at $T_{\rm eq}\approx 1$~eV, we get $\rho_\vphi(T_{\rm eq})\approx 2.6 \times 10^{-3}$~eV$^4$, well below the radiation energy density of $\ord{\text{eV}^4}$.  Hence, with the above choices for our parameters we will not overclose the Universe, and $\vphi$ will be a small subcomponent of DM.

Here, we note that we could have also used the  coupling of $\varphi$ to electrons to induce $\langle\varphi\rangle\neq 0$ in the thermal bath.  However, this contribution would be subdominant to that of the muon under generic assumptions, that is, if we assume a coupling to the muon that is not smaller than the electron's, which one may expect on general grounds.  The coupling of a long range scalar to electrons is constrained to satisfy  $g_e\lesssim 10^{-25}$ \cite{Berge:2017ovy,Fayet:2017pdp,MICROSCOPE:2022doy,Davoudiasl:2018ltz}. (Although this is a tighter constraint than the one on $g_p$, it does not give a correspondingly tighter bound on $g_\mu$ because a similar two-loop interaction is chirality suppressed \cite{Fiorillo:2026wso}.) As can be seen in \cref{fig:phi-evol}, including the electron contribution, according to \cref{vphi-evol-rel}, would only result in a negligible contribution to the shift of the $\psiB$ mass.  This is also partly due to the suppressed effect of a relativistic source for $\varphi$.  Qualitatively, this can partly be understood as follows.  A relativistically invariant fermion source for $\varphi$ can be formulated by the product of its number density $n_f$ and its inverse boost factor $\sqrt{1-v_f^2}$, where $v_f$ is a typical velocity.  Hence, electrons have a suppressed effect at high temperatures (see, for example, Ref.~\cite{Gubser:2004du}).

One may worry that the much larger coupling of $\vphi$ to $\psiB$ may lead to a back reaction and drive $\vphi$ to values larger than depicted in \cref{fig:phi-evol}.  This could push the mass of $\psiB$ to small values and hamper its decay into $\phi, \xi$ through the interactions given in \cref{psi-DM}.  However, taking $y_\psi \gg  10^{-10}$, we have assumed that the decay of $\psiB$ is prompt on the cosmological evolution time scales and a $\psiB$ would quickly decay into DM states. 

A loop generated interaction can still couple $\vphi$ to DM, for example $g_\xi \,\vphi \bar \xi \xi$.  The loop-induced coupling  can be roughly estimated by 
\beq
g_\xi \sim \frac{g_\psi y_\psi^2}{16 \pi^2}\,.
\label{eq:gxi-loop}
\eeq
One may also want to make sure that for our benchmark parameters the  induced $\vphi$ couplings to DM in \cref{eq:gxi-loop} are sufficiently small in order to avoid a large back-reaction on the misalignment of $\vphi$.  To check this, let us assume the initial decaying modulus $\Phi$, which  produced the required $B$ mesons,  reheated the Universe to $T_{\rm M}\sim 10$~MeV and that $\text{Br}(\Phi\to B \bar B)$=1.  We then roughly have 
\beq
m_B\, n_B \sim T_M^4 \Rightarrow
n_B\sim 10^{-3}\, T_M^3\,. \label{eq:nB}
\eeq
Given that we want an $\ord{1}$ fraction of $B$ mesons to decay into the dark sector during mesogenesis, we infer that $n_\xi \sim 10^{-3}\, T_M^3$, initially.  

To make sure that the initial population of DM (before annihilation down to its asymmetric relic density) does not dominate the misalignment of $\vphi$ and push $m_\psi$ below $\ord{\rm GeV}$, we demand 
\beq
g_\psi\, \delta \vphi \lesssim m_\psi^0  \Rightarrow  
\frac{g_\psi\, g_\xi\, n_\xi}{m_\vphi^2} \lesssim m_\psi^0\,,
\label{eq:deltaphi}
\eeq
where $\delta \vphi$ is the contribution of DM state  $\xi$.  For $m_\vphi\sim 10^{-13}$ eV and $m_\psi^0\sim 10$~GeV, as in \cref{fig:phi-evol}, and using \cref{eq:nB,eq:gxi-loop}, we get 
\beq
g_\psi \, y_\psi \lesssim 10^{-16}\,. 
\label{eq:gpsigxi}
\eeq
Hence, for our benchmark value of $g_\psi \sim 3\times 10^{-13}$, the above relation is readily satisfied with  $y_\psi \lesssim 10^{-5}$, which is also required to relax the LHC bounds we discuss in the next section. See also the discussion below \cref{eq:Gammapsitoxiphi}.



\section{Possible Signals} \label{sec:signals}

In this section we discuss possible signals for this mechanism beyond LHC searches for color-triplet scalars. We focus instead on signals for the dark sector, such as displaced vertices,  as well as long-range muonic forces arising from the ultralight scalar. Athough we will not discuss it here, another signal can come from the physics facilitating the annihilation of the symmetric population of DM. 

\subsection{Potential collider signals}
In previous works on mesogenesis, where $\psiB$ is light, exotic decays of $B$ mesons were extensively studied. The branching fraction that can be constrained is given as \cite{Elor:2018twp} 
\begin{align}
    {\rm Br} &(B\to \psiB +{\rm baryon}) \simeq \\ \notag
    &0.5\, (y_{cb}y_{\psi s})^2 \left(\frac{m_B- m_\psi}{3~{\rm GeV}}\right)^4\left( \frac{600~{\rm GeV}}{m_\Y}\right)^4\,.
\end{align}
Belle II has the potential to probe branching fractions $ {\rm Br} (B\to \psiB +{\rm baryon})\sim 3 \times 10^{-6}$~\cite{Alonso-Alvarez:2021qfd}.  

When the mass of the $\psiB$ is too heavy to be a final state in a $B\to {\rm inv.}+ \mathcal{B}_c$ decay, there is still a possibility of the three-body $B \to \psiB^* + \mathcal{B}_c \to \xi\, \phi +\mathcal{B}_c$, where the $\psiB^*$ is off-shell. (Here $\mathcal{B}_c$ is a charm baryon.) Compared to the above estimate, this decay rate will be suppressed by  $y_\psi^2$, an extra phase space factor of $8\pi^2$ and a factor of $(m_B/m_\psi^0)^2 \lesssim 1/4$. [Note also that the mass difference $(m_B -m_\psi)$ will be replaced by $m_B - (m_\xi + m_\phi)$.] This exotic decay can be observable in the future for $y_\psi \gtrsim 10^{-2}$.  However, as we require $\psiB\to \text{jets}$ to be dominant in order to relax the LHC constraints \cite{Hiller:2026osz}, we typically need $y_\psi\lesssim 10^{-5}$, as discussed earlier, making the missing energy channel unobservable. 

Although we take $m_\psi =10~$GeV in this work as a benchmark value, masses around $7.5~{\rm GeV}\lesssim m_\psi\lesssim10~$ GeV have been identified in \cite{Hiller:2026osz} as a potential target for displaced vertex searches at the LHC. This mass regime can easily be accommodated in our scenario by only choosing  $g_\psi \sim 2\times 10^{-13}$.

\subsection{Long Range Muonic Force Probes}

For values of $g_\mu\sim 10^{-18}$, experiments searching for new long range forces can in principle be sensitive to our parameter space \cite{Berge:2017ovy,Fayet:2017pdp,MICROSCOPE:2022doy,Davoudiasl:2018ltz}. 
An intriguing possibility is that a long-range muonic force can be probed by gravitational wave signal resulting from neutron star mergers~\cite{Dror:2019uea, Liu:2025zuz}.  This is due to the presence of large numbers of muons, $N_\mu \sim 10^{55}$ (see, for example, Ref.~\cite{Garani:2019fpa}) in the dense environment of a neutron star.  It is estimated that LIGO could probe long-range muonic forces with couplings below $10^{-20}$, but a robust bound is subject to LIGO analysis.

A question related to the above setup is to what extent one can expect the mass of $\psiB$ to be lowered within a neutron star due to the significant muon abundance in the stellar environment.  This could in principle lead to proton decay, if $\psiB$ becomes lighter than a nucleon.  A simple estimate for the background value of the scalar in the star is 
\beq
\vphi_{\rm NS}\sim \frac{g_\mu\, N_\mu}{4 \pi R}\sim g_\mu 10^{37}~\text{MeV}\,,
\label{eq:vphiNS}
\eeq
where $R\sim 10$~km is a typical length scale for the neutron star. 

Since the mass of $\psiB$ is given by $m_\psi^0+g_\psi \vphi$, we see that for $m_\psi^0=10$~GeV,  we need 
\beq
g_\mu g_\psi \lesssim 10^{-33}\,,
\label{eq:gmugpsi}
\eeq
assuming $g_\mu,\, g_\psi >0$, in order to have $m_\psi \gtrsim 1$~GeV.  Therefore, our benchmark values, used in \cref{fig:phi-evol}, would not lead to the possibility of nucleon decay in a  neutron star.

\section{Conclusions}\label{sec:conc}

What is the SM \emph{CP} violation good for? Although one of the fundamental conditions for creating a baryon asymmetry is the existence of \emph{CP} violation in a theory, the effects of the CKM phase get suppressed in electroweak baryogenesis models. \emph{Mesogenesis with a morphing mediator} model~\cite{Elor:2024cea} is the only model that utilizes the observed SM \emph{CP} violation in $B$ meson systems in order to create the baryon asymmetry of the Universe.

Mesogenesis models mainly rely on effective operators given in \cref{eq:EFTop} which can arise from the UV interactions in \cref{eq:UVmodel} between SM quarks and a dark sector fermion $\psi_\B$, which carries baryon number, mediated by an $SU(3)$ triplet scalar $\Y$. On the one hand, in order for the SM \emph{CP} violation to be enough, the effective interactions need to be sufficiently large. On the other hand, large enough interactions with quarks are heavily constrained by the LHC and, specifically for mesogenesis, by $B$ factories. In order to alleviate these constraints, previous work \cite{Elor:2024cea} explored a mechanism where the mediator mass changes from $\sim\!600$ GeV to a TeV after the baryon asymmetry is created. However, this realization of mesogenesis, where the baryon asymmetry is created in certain domains with a light mediator between the SM and a dark sector, is argued to suffer from BBN constraints due to inhomogeneities~\cite{Bagherian:2025puf}.

In this work, we introduced an alternative mechanism to realize mesogenesis using still only the SM \emph{CP} violation. Instead of focusing on the mediator, we developed a model in which the dark sector fermion $\psi_\B$, in vacuum, is heavier than the $B$ meson but becomes temporarily light enough to allow for mesogenesis in the early Universe. In this scenario, since $\psi_\B$ is heavy currently,  exotic $B$ meson decay channels are shut off at colliders. This mass evolution is realized via couplings to an ultralight scalar whose initial field value is driven by the thermal number density of SM charged leptons~\cite{Croon:2020ntf}. As benchmark values, we show that a scalar with mass $\ord{10^{-13}~{\rm eV}} $ and coupling constants of $10^{-10}-10^{-24}$ are able to realize this scenario.  In this work, we focused on muon-induced thermal misalignment, but one could alternatively employ electrons, for different choices of parameters.

Here, we also included multiple terms from the mesogenesis Lagrangian in \cref{eq:UVmodel} in order to dilute the branching fraction of the mediator $\Y$ into quark and missing energy final states. Successful mesogenesis, if we insist on using only the SM $CP$ violation, requires $M_\Y\lesssim 600$ GeV. We examined several relevant LHC searches to argue that with couplings $y_{cb} \sim 1, y_{tb}\sim 5$ such a light $SU(3)$ triplet scalar could still be viable. These are very sizable couplings, giving rise to a broad resonance. It is interesting to explore further search strategies at the LHC for such a scenario, which we leave for future work.  However, we expect that the underlying physics mediator $\Y$ would be within the reach of the LHC, as more data get analyzed and accumulated.

In the benchmark scenario we considered, $m_\psi =10$~GeV, the dark fermion $\psiB$ decays to jets at the LHC. For slightly lighter $\psiB$, these decays can produce potentially detectable displaced vertex signals. The ultralight scalar can also be uncovered in terrestrial searches for long-range forces as well as neutron star mergers. 

\vskip0.5cm\emph{No data were generated during the course of this work.}

\section*{Acknowledgements}
We thank G. Hiller for alerting us to a very recent analysis \cite{Hiller:2026osz}, regarding the LHC constraints relevant to our model. The work of H.D. is supported by the US Department of Energy under Grant Contract No. DE-SC0012704. R.H. is supported by the Institute for Fundamental Theory at the University of
Florida. S.I. is
supported by the Natural Sciences and Engineering Research Council (NSERC) of Canada under Grant No. SAPIN-2022-00024. This research was supported by the Munich Institute for Astro-, Particle and BioPhysics (MIAPbP) which is funded by the Deutsche Forschungsgemeinschaft (DFG, German Research Foundation) under Germany's Excellence Strategy EXC--2094--390783311.

\twocolumngrid


\begin{thebibliography}{42}%
\makeatletter
\providecommand \@ifxundefined [1]{%
 \@ifx{#1\undefined}
}%
\providecommand \@ifnum [1]{%
 \ifnum #1\expandafter \@firstoftwo
 \else \expandafter \@secondoftwo
 \fi
}%
\providecommand \@ifx [1]{%
 \ifx #1\expandafter \@firstoftwo
 \else \expandafter \@secondoftwo
 \fi
}%
\providecommand \natexlab [1]{#1}%
\providecommand \enquote  [1]{``#1''}%
\providecommand \bibnamefont  [1]{#1}%
\providecommand \bibfnamefont [1]{#1}%
\providecommand \citenamefont [1]{#1}%
\providecommand \href@noop [0]{\@secondoftwo}%
\providecommand \href [0]{\begingroup \@sanitize@url \@href}%
\providecommand \@href[1]{\@@startlink{#1}\@@href}%
\providecommand \@@href[1]{\endgroup#1\@@endlink}%
\providecommand \@sanitize@url [0]{\catcode `\\12\catcode `\$12\catcode
  `\&12\catcode `\#12\catcode `\^12\catcode `\_12\catcode `\%12\relax}%
\providecommand \@@startlink[1]{}%
\providecommand \@@endlink[0]{}%
\providecommand \url  [0]{\begingroup\@sanitize@url \@url }%
\providecommand \@url [1]{\endgroup\@href {#1}{\urlprefix }}%
\providecommand \urlprefix  [0]{URL }%
\providecommand \Eprint [0]{\href }%
\providecommand \doibase [0]{http://dx.doi.org/}%
\providecommand \selectlanguage [0]{\@gobble}%
\providecommand \bibinfo  [0]{\@secondoftwo}%
\providecommand \bibfield  [0]{\@secondoftwo}%
\providecommand \translation [1]{[#1]}%
\providecommand \BibitemOpen [0]{}%
\providecommand \bibitemStop [0]{}%
\providecommand \bibitemNoStop [0]{.\EOS\space}%
\providecommand \EOS [0]{\spacefactor3000\relax}%
\providecommand \BibitemShut  [1]{\csname bibitem#1\endcsname}%
\let\auto@bib@innerbib\@empty
\bibitem [{\citenamefont {Tanabashi}\ \emph {et~al.}(2018)\citenamefont
  {Tanabashi} \emph {et~al.}}]{pdg}%
  \BibitemOpen
  \bibfield  {author} {\bibinfo {author} {\bibfnamefont {M.}~\bibnamefont
  {Tanabashi}} \emph {et~al.} (\bibinfo {collaboration} {ParticleDataGroup}),\
  }\href {\doibase 10.1103/PhysRevD.98.030001} {\bibfield  {journal} {\bibinfo
  {journal} {Phys. Rev.}\ }\textbf {\bibinfo {volume} {D98}},\ \bibinfo {pages}
  {030001} (\bibinfo {year} {2018})}\BibitemShut {NoStop}%
\bibitem [{\citenamefont {Sakharov}(1967)}]{Sakharov:1967dj}%
  \BibitemOpen
  \bibfield  {author} {\bibinfo {author} {\bibfnamefont {A.~D.}\ \bibnamefont
  {Sakharov}},\ }\href {\doibase 10.1070/PU1991v034n05ABEH002497} {\bibfield
  {journal} {\bibinfo  {journal} {Pisma Zh. Eksp. Teor. Fiz.}\ }\textbf
  {\bibinfo {volume} {5}},\ \bibinfo {pages} {32} (\bibinfo {year}
  {1967})}\BibitemShut {NoStop}%
\bibitem [{\citenamefont {Gavela}\ \emph
  {et~al.}(1994{\natexlab{a}})\citenamefont {Gavela}, \citenamefont
  {Hernandez}, \citenamefont {Orloff},\ and\ \citenamefont
  {Pene}}]{Gavela:1993ts}%
  \BibitemOpen
  \bibfield  {author} {\bibinfo {author} {\bibfnamefont {M.~B.}\ \bibnamefont
  {Gavela}}, \bibinfo {author} {\bibfnamefont {P.}~\bibnamefont {Hernandez}},
  \bibinfo {author} {\bibfnamefont {J.}~\bibnamefont {Orloff}}, \ and\ \bibinfo
  {author} {\bibfnamefont {O.}~\bibnamefont {Pene}},\ }\href {\doibase
  10.1142/S0217732394000629} {\bibfield  {journal} {\bibinfo  {journal} {Mod.
  Phys. Lett. A}\ }\textbf {\bibinfo {volume} {9}},\ \bibinfo {pages} {795}
  (\bibinfo {year} {1994}{\natexlab{a}})},\ \Eprint
  {http://arxiv.org/abs/hep-ph/9312215} {arXiv:hep-ph/9312215} \BibitemShut
  {NoStop}%
\bibitem [{\citenamefont {Gavela}\ \emph
  {et~al.}(1994{\natexlab{b}})\citenamefont {Gavela}, \citenamefont {Lozano},
  \citenamefont {Orloff},\ and\ \citenamefont {Pene}}]{Gavela:1994ds}%
  \BibitemOpen
  \bibfield  {author} {\bibinfo {author} {\bibfnamefont {M.~B.}\ \bibnamefont
  {Gavela}}, \bibinfo {author} {\bibfnamefont {M.}~\bibnamefont {Lozano}},
  \bibinfo {author} {\bibfnamefont {J.}~\bibnamefont {Orloff}}, \ and\ \bibinfo
  {author} {\bibfnamefont {O.}~\bibnamefont {Pene}},\ }\href {\doibase
  10.1016/0550-3213(94)00409-9} {\bibfield  {journal} {\bibinfo  {journal}
  {Nucl. Phys. B}\ }\textbf {\bibinfo {volume} {430}},\ \bibinfo {pages} {345}
  (\bibinfo {year} {1994}{\natexlab{b}})},\ \Eprint
  {http://arxiv.org/abs/hep-ph/9406288} {arXiv:hep-ph/9406288} \BibitemShut
  {NoStop}%
\bibitem [{\citenamefont {Gavela}\ \emph
  {et~al.}(1994{\natexlab{c}})\citenamefont {Gavela}, \citenamefont
  {Hernandez}, \citenamefont {Orloff}, \citenamefont {Pene},\ and\
  \citenamefont {Quimbay}}]{Gavela:1994dt}%
  \BibitemOpen
  \bibfield  {author} {\bibinfo {author} {\bibfnamefont {M.~B.}\ \bibnamefont
  {Gavela}}, \bibinfo {author} {\bibfnamefont {P.}~\bibnamefont {Hernandez}},
  \bibinfo {author} {\bibfnamefont {J.}~\bibnamefont {Orloff}}, \bibinfo
  {author} {\bibfnamefont {O.}~\bibnamefont {Pene}}, \ and\ \bibinfo {author}
  {\bibfnamefont {C.}~\bibnamefont {Quimbay}},\ }\href {\doibase
  10.1016/0550-3213(94)00410-2} {\bibfield  {journal} {\bibinfo  {journal}
  {Nucl. Phys. B}\ }\textbf {\bibinfo {volume} {430}},\ \bibinfo {pages} {382}
  (\bibinfo {year} {1994}{\natexlab{c}})},\ \Eprint
  {http://arxiv.org/abs/hep-ph/9406289} {arXiv:hep-ph/9406289} \BibitemShut
  {NoStop}%
\bibitem [{\citenamefont {Huet}\ and\ \citenamefont
  {Sather}(1995)}]{Huet:1994jb}%
  \BibitemOpen
  \bibfield  {author} {\bibinfo {author} {\bibfnamefont {P.}~\bibnamefont
  {Huet}}\ and\ \bibinfo {author} {\bibfnamefont {E.}~\bibnamefont {Sather}},\
  }\href {\doibase 10.1103/PhysRevD.51.379} {\bibfield  {journal} {\bibinfo
  {journal} {Phys. Rev. D}\ }\textbf {\bibinfo {volume} {51}},\ \bibinfo
  {pages} {379} (\bibinfo {year} {1995})},\ \Eprint
  {http://arxiv.org/abs/hep-ph/9404302} {arXiv:hep-ph/9404302} \BibitemShut
  {NoStop}%
\bibitem [{\citenamefont {Elor}\ \emph {et~al.}(2025)\citenamefont {Elor},
  \citenamefont {Houtz}, \citenamefont {Ipek},\ and\ \citenamefont
  {Ulloa}}]{Elor:2024cea}%
  \BibitemOpen
  \bibfield  {author} {\bibinfo {author} {\bibfnamefont {G.}~\bibnamefont
  {Elor}}, \bibinfo {author} {\bibfnamefont {R.}~\bibnamefont {Houtz}},
  \bibinfo {author} {\bibfnamefont {S.}~\bibnamefont {Ipek}}, \ and\ \bibinfo
  {author} {\bibfnamefont {M.}~\bibnamefont {Ulloa}},\ }\href {\doibase
  10.1103/glgl-v4v6} {\bibfield  {journal} {\bibinfo  {journal} {Phys. Rev. D}\
  }\textbf {\bibinfo {volume} {112}},\ \bibinfo {pages} {L011701} (\bibinfo
  {year} {2025})},\ \Eprint {http://arxiv.org/abs/2408.12647} {arXiv:2408.12647
  [hep-ph]} \BibitemShut {NoStop}%
\bibitem [{\citenamefont {Berkooz}\ \emph {et~al.}(2004)\citenamefont
  {Berkooz}, \citenamefont {Nir},\ and\ \citenamefont
  {Volansky}}]{Berkooz:2004kx}%
  \BibitemOpen
  \bibfield  {author} {\bibinfo {author} {\bibfnamefont {M.}~\bibnamefont
  {Berkooz}}, \bibinfo {author} {\bibfnamefont {Y.}~\bibnamefont {Nir}}, \ and\
  \bibinfo {author} {\bibfnamefont {T.}~\bibnamefont {Volansky}},\ }\href
  {\doibase 10.1103/PhysRevLett.93.051301} {\bibfield  {journal} {\bibinfo
  {journal} {Phys. Rev. Lett.}\ }\textbf {\bibinfo {volume} {93}},\ \bibinfo
  {pages} {051301} (\bibinfo {year} {2004})},\ \Eprint
  {http://arxiv.org/abs/hep-ph/0401012} {arXiv:hep-ph/0401012} \BibitemShut
  {NoStop}%
\bibitem [{\citenamefont {Perez}\ and\ \citenamefont
  {Volansky}(2005)}]{Perez:2005yx}%
  \BibitemOpen
  \bibfield  {author} {\bibinfo {author} {\bibfnamefont {G.}~\bibnamefont
  {Perez}}\ and\ \bibinfo {author} {\bibfnamefont {T.}~\bibnamefont
  {Volansky}},\ }\href {\doibase 10.1103/PhysRevD.72.103522} {\bibfield
  {journal} {\bibinfo  {journal} {Phys. Rev. D}\ }\textbf {\bibinfo {volume}
  {72}},\ \bibinfo {pages} {103522} (\bibinfo {year} {2005})},\ \Eprint
  {http://arxiv.org/abs/hep-ph/0505222} {arXiv:hep-ph/0505222} \BibitemShut
  {NoStop}%
\bibitem [{\citenamefont {Bruggisser}\ \emph {et~al.}(2017)\citenamefont
  {Bruggisser}, \citenamefont {Konstandin},\ and\ \citenamefont
  {Servant}}]{Bruggisser:2017lhc}%
  \BibitemOpen
  \bibfield  {author} {\bibinfo {author} {\bibfnamefont {S.}~\bibnamefont
  {Bruggisser}}, \bibinfo {author} {\bibfnamefont {T.}~\bibnamefont
  {Konstandin}}, \ and\ \bibinfo {author} {\bibfnamefont {G.}~\bibnamefont
  {Servant}},\ }\href {\doibase 10.1088/1475-7516/2017/11/034} {\bibfield
  {journal} {\bibinfo  {journal} {JCAP}\ }\textbf {\bibinfo {volume} {11}},\
  \bibinfo {pages} {034} (\bibinfo {year} {2017})},\ \Eprint
  {http://arxiv.org/abs/1706.08534} {arXiv:1706.08534 [hep-ph]} \BibitemShut
  {NoStop}%
\bibitem [{\citenamefont {Baldes}\ \emph {et~al.}(2016)\citenamefont {Baldes},
  \citenamefont {Konstandin},\ and\ \citenamefont {Servant}}]{Baldes_2016}%
  \BibitemOpen
  \bibfield  {author} {\bibinfo {author} {\bibfnamefont {I.}~\bibnamefont
  {Baldes}}, \bibinfo {author} {\bibfnamefont {T.}~\bibnamefont {Konstandin}},
  \ and\ \bibinfo {author} {\bibfnamefont {G.}~\bibnamefont {Servant}},\ }\href
  {\doibase 10.1007/jhep12(2016)073} {\bibfield  {journal} {\bibinfo  {journal}
  {Journal of High Energy Physics}\ }\textbf {\bibinfo {volume} {2016}}
  (\bibinfo {year} {2016}),\ 10.1007/jhep12(2016)073}\BibitemShut {NoStop}%
\bibitem [{\citenamefont {von Harling}\ and\ \citenamefont
  {Servant}(2017)}]{von_Harling_2017}%
  \BibitemOpen
  \bibfield  {author} {\bibinfo {author} {\bibfnamefont {B.}~\bibnamefont {von
  Harling}}\ and\ \bibinfo {author} {\bibfnamefont {G.}~\bibnamefont
  {Servant}},\ }\href {\doibase 10.1007/jhep05(2017)077} {\bibfield  {journal}
  {\bibinfo  {journal} {Journal of High Energy Physics}\ }\textbf {\bibinfo
  {volume} {2017}} (\bibinfo {year} {2017}),\
  10.1007/jhep05(2017)077}\BibitemShut {NoStop}%
\bibitem [{\citenamefont {Baldes}\ \emph {et~al.}(2018)\citenamefont {Baldes},
  \citenamefont {Konstandin},\ and\ \citenamefont {Servant}}]{Baldes_2018}%
  \BibitemOpen
  \bibfield  {author} {\bibinfo {author} {\bibfnamefont {I.}~\bibnamefont
  {Baldes}}, \bibinfo {author} {\bibfnamefont {T.}~\bibnamefont {Konstandin}},
  \ and\ \bibinfo {author} {\bibfnamefont {G.}~\bibnamefont {Servant}},\ }\href
  {\doibase 10.1016/j.physletb.2018.10.015} {\bibfield  {journal} {\bibinfo
  {journal} {Physics Letters B}\ }\textbf {\bibinfo {volume} {786}},\ \bibinfo
  {pages} {373–377} (\bibinfo {year} {2018})}\BibitemShut {NoStop}%
\bibitem [{\citenamefont {Bruggisser}\ \emph
  {et~al.}(2018{\natexlab{a}})\citenamefont {Bruggisser}, \citenamefont {von
  Harling}, \citenamefont {Matsedonskyi},\ and\ \citenamefont
  {Servant}}]{Bruggisser_2018}%
  \BibitemOpen
  \bibfield  {author} {\bibinfo {author} {\bibfnamefont {S.}~\bibnamefont
  {Bruggisser}}, \bibinfo {author} {\bibfnamefont {B.}~\bibnamefont {von
  Harling}}, \bibinfo {author} {\bibfnamefont {O.}~\bibnamefont
  {Matsedonskyi}}, \ and\ \bibinfo {author} {\bibfnamefont {G.}~\bibnamefont
  {Servant}},\ }\href {\doibase 10.1103/physrevlett.121.131801} {\bibfield
  {journal} {\bibinfo  {journal} {Physical Review Letters}\ }\textbf {\bibinfo
  {volume} {121}} (\bibinfo {year} {2018}{\natexlab{a}}),\
  10.1103/physrevlett.121.131801}\BibitemShut {NoStop}%
\bibitem [{\citenamefont {Bruggisser}\ \emph
  {et~al.}(2018{\natexlab{b}})\citenamefont {Bruggisser}, \citenamefont {von
  Harling}, \citenamefont {Matsedonskyi},\ and\ \citenamefont
  {Servant}}]{bruggisser2018electroweakphasetransitionbaryogenesis}%
  \BibitemOpen
  \bibfield  {author} {\bibinfo {author} {\bibfnamefont {S.}~\bibnamefont
  {Bruggisser}}, \bibinfo {author} {\bibfnamefont {B.}~\bibnamefont {von
  Harling}}, \bibinfo {author} {\bibfnamefont {O.}~\bibnamefont
  {Matsedonskyi}}, \ and\ \bibinfo {author} {\bibfnamefont {G.}~\bibnamefont
  {Servant}},\ }\href {https://arxiv.org/abs/1804.07314} {\enquote {\bibinfo
  {title} {Electroweak phase transition and baryogenesis in composite higgs
  models},}\ } (\bibinfo {year} {2018}{\natexlab{b}}),\ \Eprint
  {http://arxiv.org/abs/1804.07314} {arXiv:1804.07314 [hep-ph]} \BibitemShut
  {NoStop}%
\bibitem [{\citenamefont {Fuyuto}\ \emph {et~al.}(2018)\citenamefont {Fuyuto},
  \citenamefont {Hou},\ and\ \citenamefont {Senaha}}]{Fuyuto:2017ewj}%
  \BibitemOpen
  \bibfield  {author} {\bibinfo {author} {\bibfnamefont {K.}~\bibnamefont
  {Fuyuto}}, \bibinfo {author} {\bibfnamefont {W.-S.}\ \bibnamefont {Hou}}, \
  and\ \bibinfo {author} {\bibfnamefont {E.}~\bibnamefont {Senaha}},\ }\href
  {\doibase 10.1016/j.physletb.2017.11.073} {\bibfield  {journal} {\bibinfo
  {journal} {Phys. Lett. B}\ }\textbf {\bibinfo {volume} {776}},\ \bibinfo
  {pages} {402} (\bibinfo {year} {2018})},\ \Eprint
  {http://arxiv.org/abs/1705.05034} {arXiv:1705.05034 [hep-ph]} \BibitemShut
  {NoStop}%
\bibitem [{\citenamefont {Elahi}\ \emph {et~al.}(2022)\citenamefont {Elahi},
  \citenamefont {Elor},\ and\ \citenamefont {McGehee}}]{Elahi:2021jia}%
  \BibitemOpen
  \bibfield  {author} {\bibinfo {author} {\bibfnamefont {F.}~\bibnamefont
  {Elahi}}, \bibinfo {author} {\bibfnamefont {G.}~\bibnamefont {Elor}}, \ and\
  \bibinfo {author} {\bibfnamefont {R.}~\bibnamefont {McGehee}},\ }\href
  {\doibase 10.1103/PhysRevD.105.055024} {\bibfield  {journal} {\bibinfo
  {journal} {Phys. Rev. D}\ }\textbf {\bibinfo {volume} {105}},\ \bibinfo
  {pages} {055024} (\bibinfo {year} {2022})},\ \Eprint
  {http://arxiv.org/abs/2109.09751} {arXiv:2109.09751 [hep-ph]} \BibitemShut
  {NoStop}%
\bibitem [{\citenamefont {Elor}\ \emph {et~al.}(2019)\citenamefont {Elor},
  \citenamefont {Escudero},\ and\ \citenamefont {Nelson}}]{Elor:2018twp}%
  \BibitemOpen
  \bibfield  {author} {\bibinfo {author} {\bibfnamefont {G.}~\bibnamefont
  {Elor}}, \bibinfo {author} {\bibfnamefont {M.}~\bibnamefont {Escudero}}, \
  and\ \bibinfo {author} {\bibfnamefont {A.}~\bibnamefont {Nelson}},\ }\href
  {\doibase 10.1103/PhysRevD.99.035031} {\bibfield  {journal} {\bibinfo
  {journal} {Phys. Rev. D}\ }\textbf {\bibinfo {volume} {99}},\ \bibinfo
  {pages} {035031} (\bibinfo {year} {2019})},\ \Eprint
  {http://arxiv.org/abs/1810.00880} {arXiv:1810.00880 [hep-ph]} \BibitemShut
  {NoStop}%
\bibitem [{\citenamefont {Alonso-\'Alvarez}\ \emph {et~al.}(2020)\citenamefont
  {Alonso-\'Alvarez}, \citenamefont {Elor}, \citenamefont {Nelson},\ and\
  \citenamefont {Xiao}}]{Alonso-Alvarez:2019fym}%
  \BibitemOpen
  \bibfield  {author} {\bibinfo {author} {\bibfnamefont {G.}~\bibnamefont
  {Alonso-\'Alvarez}}, \bibinfo {author} {\bibfnamefont {G.}~\bibnamefont
  {Elor}}, \bibinfo {author} {\bibfnamefont {A.~E.}\ \bibnamefont {Nelson}}, \
  and\ \bibinfo {author} {\bibfnamefont {H.}~\bibnamefont {Xiao}},\ }\href
  {\doibase 10.1007/JHEP03(2020)046} {\bibfield  {journal} {\bibinfo  {journal}
  {JHEP}\ }\textbf {\bibinfo {volume} {03}},\ \bibinfo {pages} {046} (\bibinfo
  {year} {2020})},\ \Eprint {http://arxiv.org/abs/1907.10612} {arXiv:1907.10612
  [hep-ph]} \BibitemShut {NoStop}%
\bibitem [{\citenamefont {Nelson}\ and\ \citenamefont
  {Xiao}(2019)}]{Nelson:2019fln}%
  \BibitemOpen
  \bibfield  {author} {\bibinfo {author} {\bibfnamefont {A.~E.}\ \bibnamefont
  {Nelson}}\ and\ \bibinfo {author} {\bibfnamefont {H.}~\bibnamefont {Xiao}},\
  }\href@noop {} {\  (\bibinfo {year} {2019})},\ \Eprint
  {http://arxiv.org/abs/1901.08141} {arXiv:1901.08141 [hep-ph]} \BibitemShut
  {NoStop}%
\bibitem [{\citenamefont {Bagherian}\ \emph {et~al.}(2026)\citenamefont
  {Bagherian}, \citenamefont {Ekhterachian},\ and\ \citenamefont
  {Stelzl}}]{Bagherian:2025puf}%
  \BibitemOpen
  \bibfield  {author} {\bibinfo {author} {\bibfnamefont {H.}~\bibnamefont
  {Bagherian}}, \bibinfo {author} {\bibfnamefont {M.}~\bibnamefont
  {Ekhterachian}}, \ and\ \bibinfo {author} {\bibfnamefont {S.}~\bibnamefont
  {Stelzl}},\ }\href {\doibase 10.1007/JHEP01(2026)068} {\bibfield  {journal}
  {\bibinfo  {journal} {JHEP}\ }\textbf {\bibinfo {volume} {01}},\ \bibinfo
  {pages} {068} (\bibinfo {year} {2026})},\ \Eprint
  {http://arxiv.org/abs/2505.15904} {arXiv:2505.15904 [hep-ph]} \BibitemShut
  {NoStop}%
\bibitem [{\citenamefont {Baruch}\ \emph {et~al.}(2026)\citenamefont {Baruch},
  \citenamefont {Elor}, \citenamefont {Goldberg}, \citenamefont {Shtaif},\ and\
  \citenamefont {Soreq}}]{Baruch:2026iwn}%
  \BibitemOpen
  \bibfield  {author} {\bibinfo {author} {\bibfnamefont {C.}~\bibnamefont
  {Baruch}}, \bibinfo {author} {\bibfnamefont {G.}~\bibnamefont {Elor}},
  \bibinfo {author} {\bibfnamefont {J.~M.}\ \bibnamefont {Goldberg}}, \bibinfo
  {author} {\bibfnamefont {O.}~\bibnamefont {Shtaif}}, \ and\ \bibinfo {author}
  {\bibfnamefont {Y.}~\bibnamefont {Soreq}},\ }\href@noop {} {\  (\bibinfo
  {year} {2026})},\ \Eprint {http://arxiv.org/abs/2603.28330} {arXiv:2603.28330
  [hep-ph]} \BibitemShut {NoStop}%
\bibitem [{\citenamefont {Batell}\ and\ \citenamefont
  {Ghalsasi}(2023)}]{Batell:2021ofv}%
  \BibitemOpen
  \bibfield  {author} {\bibinfo {author} {\bibfnamefont {B.}~\bibnamefont
  {Batell}}\ and\ \bibinfo {author} {\bibfnamefont {A.}~\bibnamefont
  {Ghalsasi}},\ }\href {\doibase 10.1103/PhysRevD.107.L091701} {\bibfield
  {journal} {\bibinfo  {journal} {Phys. Rev. D}\ }\textbf {\bibinfo {volume}
  {107}},\ \bibinfo {pages} {L091701} (\bibinfo {year} {2023})},\ \Eprint
  {http://arxiv.org/abs/2109.04476} {arXiv:2109.04476 [hep-ph]} \BibitemShut
  {NoStop}%
\bibitem [{\citenamefont {Croon}\ \emph
  {et~al.}(2022{\natexlab{a}})\citenamefont {Croon}, \citenamefont {Elor},
  \citenamefont {Houtz}, \citenamefont {Murayama},\ and\ \citenamefont
  {White}}]{Croon:2020ntf}%
  \BibitemOpen
  \bibfield  {author} {\bibinfo {author} {\bibfnamefont {D.}~\bibnamefont
  {Croon}}, \bibinfo {author} {\bibfnamefont {G.}~\bibnamefont {Elor}},
  \bibinfo {author} {\bibfnamefont {R.}~\bibnamefont {Houtz}}, \bibinfo
  {author} {\bibfnamefont {H.}~\bibnamefont {Murayama}}, \ and\ \bibinfo
  {author} {\bibfnamefont {G.}~\bibnamefont {White}},\ }\href {\doibase
  10.1103/PhysRevD.105.L061303} {\bibfield  {journal} {\bibinfo  {journal}
  {Phys. Rev. D}\ }\textbf {\bibinfo {volume} {105}},\ \bibinfo {pages}
  {L061303} (\bibinfo {year} {2022}{\natexlab{a}})},\ \Eprint
  {http://arxiv.org/abs/2012.15284} {arXiv:2012.15284 [hep-ph]} \BibitemShut
  {NoStop}%
\bibitem [{\citenamefont {Lenz}\ and\ \citenamefont
  {Tetlalmatzi-Xolocotzi}(2020)}]{Lenz:2019lvd}%
  \BibitemOpen
  \bibfield  {author} {\bibinfo {author} {\bibfnamefont {A.}~\bibnamefont
  {Lenz}}\ and\ \bibinfo {author} {\bibfnamefont {G.}~\bibnamefont
  {Tetlalmatzi-Xolocotzi}},\ }\href {\doibase 10.1007/JHEP07(2020)177}
  {\bibfield  {journal} {\bibinfo  {journal} {JHEP}\ }\textbf {\bibinfo
  {volume} {07}},\ \bibinfo {pages} {177} (\bibinfo {year} {2020})},\ \Eprint
  {http://arxiv.org/abs/1912.07621} {arXiv:1912.07621 [hep-ph]} \BibitemShut
  {NoStop}%
\bibitem [{\citenamefont {Alonso-\'Alvarez}\ \emph {et~al.}(2021)\citenamefont
  {Alonso-\'Alvarez}, \citenamefont {Elor},\ and\ \citenamefont
  {Escudero}}]{Alonso-Alvarez:2021qfd}%
  \BibitemOpen
  \bibfield  {author} {\bibinfo {author} {\bibfnamefont {G.}~\bibnamefont
  {Alonso-\'Alvarez}}, \bibinfo {author} {\bibfnamefont {G.}~\bibnamefont
  {Elor}}, \ and\ \bibinfo {author} {\bibfnamefont {M.}~\bibnamefont
  {Escudero}},\ }\href {\doibase 10.1103/PhysRevD.104.035028} {\bibfield
  {journal} {\bibinfo  {journal} {Phys. Rev. D}\ }\textbf {\bibinfo {volume}
  {104}},\ \bibinfo {pages} {035028} (\bibinfo {year} {2021})},\ \Eprint
  {http://arxiv.org/abs/2101.02706} {arXiv:2101.02706 [hep-ph]} \BibitemShut
  {NoStop}%
\bibitem [{\citenamefont {Hadjivasiliou}\ \emph {et~al.}(2022)\citenamefont
  {Hadjivasiliou} \emph {et~al.}}]{Belle:2021gmc}%
  \BibitemOpen
  \bibfield  {author} {\bibinfo {author} {\bibfnamefont {C.}~\bibnamefont
  {Hadjivasiliou}} \emph {et~al.} (\bibinfo {collaboration} {Belle}),\ }\href
  {\doibase 10.1103/PhysRevD.105.L051101} {\bibfield  {journal} {\bibinfo
  {journal} {Phys. Rev. D}\ }\textbf {\bibinfo {volume} {105}},\ \bibinfo
  {pages} {L051101} (\bibinfo {year} {2022})},\ \Eprint
  {http://arxiv.org/abs/2110.14086} {arXiv:2110.14086 [hep-ex]} \BibitemShut
  {NoStop}%
\bibitem [{\citenamefont {Lees}\ \emph {et~al.}(2023)\citenamefont {Lees} \emph
  {et~al.}}]{BaBar:2023rer}%
  \BibitemOpen
  \bibfield  {author} {\bibinfo {author} {\bibfnamefont {J.~P.}\ \bibnamefont
  {Lees}} \emph {et~al.} (\bibinfo {collaboration} {BaBar}),\ }\href {\doibase
  10.1103/PhysRevD.107.092001} {\bibfield  {journal} {\bibinfo  {journal}
  {Phys. Rev. D}\ }\textbf {\bibinfo {volume} {107}},\ \bibinfo {pages}
  {092001} (\bibinfo {year} {2023})},\ \Eprint
  {http://arxiv.org/abs/2302.00208} {arXiv:2302.00208 [hep-ex]} \BibitemShut
  {NoStop}%
\bibitem [{\citenamefont {Hiller}\ \emph {et~al.}(2026)\citenamefont {Hiller},
  \citenamefont {Rodr{\'\i}guez-S{\'a}nchez},\ and\ \citenamefont
  {Wendler}}]{Hiller:2026osz}%
  \BibitemOpen
  \bibfield  {author} {\bibinfo {author} {\bibfnamefont {G.}~\bibnamefont
  {Hiller}}, \bibinfo {author} {\bibfnamefont {A.}~\bibnamefont
  {Rodr{\'\i}guez-S{\'a}nchez}}, \ and\ \bibinfo {author} {\bibfnamefont
  {D.}~\bibnamefont {Wendler}},\ }\href@noop {} {\  (\bibinfo {year} {2026})},\
  \Eprint {http://arxiv.org/abs/2602.15936} {arXiv:2602.15936 [hep-ph]}
  \BibitemShut {NoStop}%
\bibitem [{\citenamefont {Sirunyan}\ \emph {et~al.}(2021)\citenamefont
  {Sirunyan} \emph {et~al.}}]{CMS:2021knz}%
  \BibitemOpen
  \bibfield  {author} {\bibinfo {author} {\bibfnamefont {A.~M.}\ \bibnamefont
  {Sirunyan}} \emph {et~al.} (\bibinfo {collaboration} {CMS}),\ }\href
  {\doibase 10.1103/PhysRevD.104.032006} {\bibfield  {journal} {\bibinfo
  {journal} {Phys. Rev. D}\ }\textbf {\bibinfo {volume} {104}},\ \bibinfo
  {pages} {032006} (\bibinfo {year} {2021})},\ \Eprint
  {http://arxiv.org/abs/2102.06976} {arXiv:2102.06976 [hep-ex]} \BibitemShut
  {NoStop}%
\bibitem [{\citenamefont {Choudhury}\ \emph {et~al.}(2024)\citenamefont
  {Choudhury}, \citenamefont {Mondal},\ and\ \citenamefont
  {Mondal}}]{Choudhury:2024ggy}%
  \BibitemOpen
  \bibfield  {author} {\bibinfo {author} {\bibfnamefont {A.}~\bibnamefont
  {Choudhury}}, \bibinfo {author} {\bibfnamefont {A.}~\bibnamefont {Mondal}}, \
  and\ \bibinfo {author} {\bibfnamefont {S.}~\bibnamefont {Mondal}},\ }\href
  {\doibase 10.1140/epjs/s11734-024-01100-x} {\  (\bibinfo {year} {2024}),\
  10.1140/epjs/s11734-024-01100-x},\ \Eprint {http://arxiv.org/abs/2402.04040}
  {arXiv:2402.04040 [hep-ph]} \BibitemShut {NoStop}%
\bibitem [{\citenamefont {Sirunyan}\ \emph {et~al.}(2018)\citenamefont
  {Sirunyan} \emph {et~al.}}]{CMS:2018mgb}%
  \BibitemOpen
  \bibfield  {author} {\bibinfo {author} {\bibfnamefont {A.~M.}\ \bibnamefont
  {Sirunyan}} \emph {et~al.} (\bibinfo {collaboration} {CMS}),\ }\href
  {\doibase 10.1007/JHEP08(2018)130} {\bibfield  {journal} {\bibinfo  {journal}
  {JHEP}\ }\textbf {\bibinfo {volume} {08}},\ \bibinfo {pages} {130} (\bibinfo
  {year} {2018})},\ \Eprint {http://arxiv.org/abs/1806.00843} {arXiv:1806.00843
  [hep-ex]} \BibitemShut {NoStop}%
\bibitem [{\citenamefont {Croon}\ \emph
  {et~al.}(2022{\natexlab{b}})\citenamefont {Croon}, \citenamefont
  {Davoudiasl},\ and\ \citenamefont {Houtz}}]{Croon:2022gwq}%
  \BibitemOpen
  \bibfield  {author} {\bibinfo {author} {\bibfnamefont {D.}~\bibnamefont
  {Croon}}, \bibinfo {author} {\bibfnamefont {H.}~\bibnamefont {Davoudiasl}}, \
  and\ \bibinfo {author} {\bibfnamefont {R.}~\bibnamefont {Houtz}},\ }\href
  {\doibase 10.1103/PhysRevD.106.035006} {\bibfield  {journal} {\bibinfo
  {journal} {Phys. Rev. D}\ }\textbf {\bibinfo {volume} {106}},\ \bibinfo
  {pages} {035006} (\bibinfo {year} {2022}{\natexlab{b}})},\ \Eprint
  {http://arxiv.org/abs/2204.07584} {arXiv:2204.07584 [hep-ph]} \BibitemShut
  {NoStop}%
\bibitem [{\citenamefont {Fiorillo}\ \emph {et~al.}(2026)\citenamefont
  {Fiorillo}, \citenamefont {Lella}, \citenamefont {Raffelt}, \citenamefont
  {Selimovic},\ and\ \citenamefont {Vitagliano}}]{Fiorillo:2026wso}%
  \BibitemOpen
  \bibfield  {author} {\bibinfo {author} {\bibfnamefont {D.~F.~G.}\
  \bibnamefont {Fiorillo}}, \bibinfo {author} {\bibfnamefont {A.}~\bibnamefont
  {Lella}}, \bibinfo {author} {\bibfnamefont {G.~G.}\ \bibnamefont {Raffelt}},
  \bibinfo {author} {\bibfnamefont {N.}~\bibnamefont {Selimovic}}, \ and\
  \bibinfo {author} {\bibfnamefont {E.}~\bibnamefont {Vitagliano}},\
  }\href@noop {} {\  (\bibinfo {year} {2026})},\ \Eprint
  {http://arxiv.org/abs/2605.24094} {arXiv:2605.24094 [hep-ph]} \BibitemShut
  {NoStop}%
\bibitem [{\citenamefont {Fayet}(2018)}]{Fayet:2017pdp}%
  \BibitemOpen
  \bibfield  {author} {\bibinfo {author} {\bibfnamefont {P.}~\bibnamefont
  {Fayet}},\ }\href {\doibase 10.1103/PhysRevD.97.055039} {\bibfield  {journal}
  {\bibinfo  {journal} {Phys. Rev. D}\ }\textbf {\bibinfo {volume} {97}},\
  \bibinfo {pages} {055039} (\bibinfo {year} {2018})},\ \Eprint
  {http://arxiv.org/abs/1712.00856} {arXiv:1712.00856 [hep-ph]} \BibitemShut
  {NoStop}%
\bibitem [{\citenamefont {Touboul}\ \emph {et~al.}(2022)\citenamefont {Touboul}
  \emph {et~al.}}]{MICROSCOPE:2022doy}%
  \BibitemOpen
  \bibfield  {author} {\bibinfo {author} {\bibfnamefont {P.}~\bibnamefont
  {Touboul}} \emph {et~al.} (\bibinfo {collaboration} {MICROSCOPE}),\ }\href
  {\doibase 10.1103/PhysRevLett.129.121102} {\bibfield  {journal} {\bibinfo
  {journal} {Phys. Rev. Lett.}\ }\textbf {\bibinfo {volume} {129}},\ \bibinfo
  {pages} {121102} (\bibinfo {year} {2022})},\ \Eprint
  {http://arxiv.org/abs/2209.15487} {arXiv:2209.15487 [gr-qc]} \BibitemShut
  {NoStop}%
\bibitem [{\citenamefont {Berg{\'e}}\ \emph {et~al.}(2018)\citenamefont
  {Berg{\'e}}, \citenamefont {Brax}, \citenamefont {M{\'e}tris}, \citenamefont
  {Pernot-Borr{\`a}s}, \citenamefont {Touboul},\ and\ \citenamefont
  {Uzan}}]{Berge:2017ovy}%
  \BibitemOpen
  \bibfield  {author} {\bibinfo {author} {\bibfnamefont {J.}~\bibnamefont
  {Berg{\'e}}}, \bibinfo {author} {\bibfnamefont {P.}~\bibnamefont {Brax}},
  \bibinfo {author} {\bibfnamefont {G.}~\bibnamefont {M{\'e}tris}}, \bibinfo
  {author} {\bibfnamefont {M.}~\bibnamefont {Pernot-Borr{\`a}s}}, \bibinfo
  {author} {\bibfnamefont {P.}~\bibnamefont {Touboul}}, \ and\ \bibinfo
  {author} {\bibfnamefont {J.-P.}\ \bibnamefont {Uzan}},\ }\href {\doibase
  10.1103/PhysRevLett.120.141101} {\bibfield  {journal} {\bibinfo  {journal}
  {Phys. Rev. Lett.}\ }\textbf {\bibinfo {volume} {120}},\ \bibinfo {pages}
  {141101} (\bibinfo {year} {2018})},\ \Eprint
  {http://arxiv.org/abs/1712.00483} {arXiv:1712.00483 [gr-qc]} \BibitemShut
  {NoStop}%
\bibitem [{\citenamefont {Davoudiasl}\ and\ \citenamefont
  {Giardino}(2019)}]{Davoudiasl:2018ltz}%
  \BibitemOpen
  \bibfield  {author} {\bibinfo {author} {\bibfnamefont {H.}~\bibnamefont
  {Davoudiasl}}\ and\ \bibinfo {author} {\bibfnamefont {P.~P.}\ \bibnamefont
  {Giardino}},\ }\href {\doibase 10.1016/j.physletb.2018.11.041} {\bibfield
  {journal} {\bibinfo  {journal} {Phys. Lett. B}\ }\textbf {\bibinfo {volume}
  {788}},\ \bibinfo {pages} {270} (\bibinfo {year} {2019})},\ \Eprint
  {http://arxiv.org/abs/1804.01098} {arXiv:1804.01098 [hep-ph]} \BibitemShut
  {NoStop}%
\bibitem [{\citenamefont {Gubser}\ and\ \citenamefont
  {Peebles}(2004)}]{Gubser:2004du}%
  \BibitemOpen
  \bibfield  {author} {\bibinfo {author} {\bibfnamefont {S.~S.}\ \bibnamefont
  {Gubser}}\ and\ \bibinfo {author} {\bibfnamefont {P.~J.~E.}\ \bibnamefont
  {Peebles}},\ }\href {\doibase 10.1103/PhysRevD.70.123511} {\bibfield
  {journal} {\bibinfo  {journal} {Phys. Rev. D}\ }\textbf {\bibinfo {volume}
  {70}},\ \bibinfo {pages} {123511} (\bibinfo {year} {2004})},\ \Eprint
  {http://arxiv.org/abs/hep-th/0407097} {arXiv:hep-th/0407097} \BibitemShut
  {NoStop}%
\bibitem [{\citenamefont {Dror}\ \emph {et~al.}(2020)\citenamefont {Dror},
  \citenamefont {Laha},\ and\ \citenamefont {Opferkuch}}]{Dror:2019uea}%
  \BibitemOpen
  \bibfield  {author} {\bibinfo {author} {\bibfnamefont {J.~A.}\ \bibnamefont
  {Dror}}, \bibinfo {author} {\bibfnamefont {R.}~\bibnamefont {Laha}}, \ and\
  \bibinfo {author} {\bibfnamefont {T.}~\bibnamefont {Opferkuch}},\ }\href
  {\doibase 10.1103/PhysRevD.102.023005} {\bibfield  {journal} {\bibinfo
  {journal} {Phys. Rev. D}\ }\textbf {\bibinfo {volume} {102}},\ \bibinfo
  {pages} {023005} (\bibinfo {year} {2020})},\ \Eprint
  {http://arxiv.org/abs/1909.12845} {arXiv:1909.12845 [hep-ph]} \BibitemShut
  {NoStop}%
\bibitem [{\citenamefont {Liu}\ and\ \citenamefont {Tang}(2025)}]{Liu:2025zuz}%
  \BibitemOpen
  \bibfield  {author} {\bibinfo {author} {\bibfnamefont {Z.}~\bibnamefont
  {Liu}}\ and\ \bibinfo {author} {\bibfnamefont {Z.-W.}\ \bibnamefont {Tang}},\
  }\href {\doibase 10.1103/4v46-c5zt} {\bibfield  {journal} {\bibinfo
  {journal} {Phys. Rev. D}\ }\textbf {\bibinfo {volume} {112}},\ \bibinfo
  {pages} {055021} (\bibinfo {year} {2025})},\ \Eprint
  {http://arxiv.org/abs/2501.10927} {arXiv:2501.10927 [astro-ph.HE]}
  \BibitemShut {NoStop}%
\bibitem [{\citenamefont {Garani}\ and\ \citenamefont
  {Heeck}(2019)}]{Garani:2019fpa}%
  \BibitemOpen
  \bibfield  {author} {\bibinfo {author} {\bibfnamefont {R.}~\bibnamefont
  {Garani}}\ and\ \bibinfo {author} {\bibfnamefont {J.}~\bibnamefont {Heeck}},\
  }\href {\doibase 10.1103/PhysRevD.100.035039} {\bibfield  {journal} {\bibinfo
   {journal} {Phys. Rev. D}\ }\textbf {\bibinfo {volume} {100}},\ \bibinfo
  {pages} {035039} (\bibinfo {year} {2019})},\ \Eprint
  {http://arxiv.org/abs/1906.10145} {arXiv:1906.10145 [hep-ph]} \BibitemShut
  {NoStop}%
\end{thebibliography}
\end{document}